\documentclass[11pt,a4paper]{article}
\usepackage{amsmath,amssymb}
\usepackage{epsfig,graphicx}
\usepackage{epic,eepic}
\usepackage{subfigure}
\usepackage{rotating}
\usepackage{bm}
\usepackage{axodraw4j}
\usepackage{color}
\usepackage{hyperref}
\usepackage{cite}

\newlength{\wth}
\def\Tr{{\rm Tr}}

\def\Dbarslash{\,\,{\raise.15ex\hbox{/}\mkern-12mu {\bar\D}}}
\def\Dslash{\,\,{\raise.15ex\hbox{/}\mkern-12mu \D}}
\def\delslash{\,\,{\raise.15ex\hbox{/}\mkern-9mu \partial}}
\def\delbarslash{\,\,{\raise.15ex\hbox{/}\mkern-9mu {\bar\partial}}}

 \def\L{\Lambda}

\def\D{{\cal D}}
\def\Dbarslash{\,\,{\raise.15ex\hbox{/}\mkern-12mu {\bar\D}}}
\def\delslash{\,\,{\raise.15ex\hbox{/}\mkern-9mu \partial}}
\def\Dslash{\,\,{\raise.15ex\hbox{/}\mkern-12mu \D}}

\def\={\, =\, }
\def\+{\, +\, }
\def\-{\, -\, }

\def\slashchar#1{\setbox0=\hbox{$#1$}           
\dimen0=\wd0                                 
\setbox1=\hbox{/} \dimen1=\wd1               
\ifdim\dimen0>\dimen1                        
   \rlap{\hbox to \dimen0{\hfil/\hfil}}      
   #1                                        
\else                                        
   \rlap{\hbox to \dimen1{\hfil$#1$\hfil}}   
   /                                         
\fi}

\newcommand{\be}{\begin{equation}}
\newcommand{\ee}{\end{equation}}
\def\bea{\begin{eqnarray}}
\def\eea{\end{eqnarray}}

\definecolor{saabeer}{rgb}{0,1,0}

\definecolor{durbeer}{rgb}{1,0,0}

\definecolor{durbeer2}{rgb}{0,0,0.7}

\begin{document}
\date{\mbox{ }}
\title{{\normalsize  IPPP/11/07; DCPT/11/14\hfill\mbox{}\hfill\mbox{}}\\
\vspace{2.5cm} \LARGE{\textbf{Mass Sum Rules and the Role of the Messenger Scale\\
in General Gauge Mediation}}}
\author{\Large
Joerg Jaeckel, Valentin V. Khoze and Chris Wymant\\[4ex]
\small{\em Institute for Particle Physics Phenomenology, Department of Physics,}\\
\small{\em Durham University, Durham DH1 3LE, United Kingdom}\\[2ex]
}  
\date{}
\maketitle

\vspace{3ex}

\begin{abstract}
\noindent In General Gauge Mediation (GGM), supersymmetry breaking
is communicated to the Standard Model (MSSM) sector via gauge
interactions at a high scale which we refer to as the messenger
scale. At this scale GGM predicts certain sum rules between the
MSSM sfermion masses. We investigate the validity and the ultimate
fate of these sum rules after RG evolution down to the electroweak
scale where the mass spectrum will be probed at colliders. We find
that the sum rules hold for the first two generations. However the
third generation (where sfermions are lightest) violates one of the two sum
rules by $10$ to $50 \%$ over the explored parameter space.
This constrains and quantifies the
potential use of sum rules as a signature of gauge mediation.
We also comment on the role played by the messenger scale in single- and multi-scale
GGM models.
\end{abstract}

\newpage


\section{Introduction}

Theories with gauge mediated supersymmetry breaking (see \cite{GRat} for a review)
provide a particularly simple and compelling set-up
for addressing theory and phenomenology beyond the Standard Model. One of the main appealing features of gauge
mediation (compared to gravity mediation) is its automatic avoidance of unsuppressed flavour changing interactions.
Recently there has been a surge of interest and research in the area which has lead to a significant extension and
generalisation of its original realisation, known now as ordinary gauge mediation (OGM).
Following this, the authors of \cite{GGM} have introduced the General Gauge Mediation (GGM) framework suitable for
unifying and analysing very general models of gauge mediation in a model-independent way
(the models can be weakly or strongly coupled, with explicit messenger sectors or with direct mediation, or
any combination of the above).
A detailed study of the
phenomenology of pure GGM models and their parameter spaces was presented recently in \cite{ADJKpGGM,ADJK7};
for earlier and related analyses of GGM phenomenology see also
Refs.~\cite{Carpenter:2008he,Rajaraman:2009ga,Kobayashi:2009rn,Meade:2010ji,Ruderman:2010kj}.

We now briefly summarise some general GGM results relevant for our purposes.
In the first instance, GGM is defined by the requirement that the SUSY breaking sector and the MSSM
decouple in the limit of $\alpha_{SM} \to 0$.
For gaugino and sfermion masses of MSSM this leads to the following general structure
\bea
\label{gauginosoft}
M_{\lambda_r}(M) &=&\, k_r \,\frac{\alpha_r}{4\pi}\,\,\Lambda_{G,r} \, , \\
\label{scalarsoft}
m_{\tilde{f}}^2 (M) &=&\, 2 \sum_{r=1}^3 C_2(f,r) k_r \,\frac{\alpha_r^2}{(4\pi)^2}\,\, \Lambda_{S,r}^2\, ,
\eea
where $k_r = (5/3,1,1)$, $\alpha_r (M)$ are the gauge coupling constants
and $C_2(f,r)$ are the quadratic Casimir operators of the representation $f$ under the $r^{\rm th}$ gauge group,
$C_{2}(3)=4/3$, $C_{2}(2)=3/4$, and $C_{2}(1)=Y^2$ for sfermions in the fundamental representation ($C_2=0$ for singlets of a gauge group).
Ordinary gauge mediation scenarios occupy the restricted
parameter space $\L_G\simeq\L_S$.

One important feature of these equations is that in GGM, gaugino and sfermion masses are governed by
generally independent
and unrelated mass-scales\footnote{This departure from ordinary gauge mediation is not surprising since
 Majorana gaugino masses require SUSY-breaking as well as
$R$-symmetry breaking, while the scalar masses need only the former.}, $\Lambda_{G,r}$ and $\Lambda_{S,r}$.
In addition, in general there is no requirement on the individual $\L$'s being the same for different values of
$r=1,2,3$, thus generically one expects six different $\L$ scales which determine the three gaugino masses and
the five sfermion masses for each of the three generations~\cite{GGM}.
In practice, if the messengers do not spoil unification of the gauge couplings, one should expect that the $\Lambda_{G,r}$ and
$\Lambda_{S,r}$ at the messenger scale should not strongly differ between the different $r$.

Equations \eqref{gauginosoft}, \eqref{scalarsoft} are derived at a certain high scale $M$ which
(at least in models with identifiable messengers fields)
has the meaning of the messenger masses $M_{\rm mess}$. The well-established usual approach is to take these
equations as the input values for soft SUSY breaking mass terms at high $M$ and perform the RG evolution
to determine observables in MSSM at the low electroweak scale. In the context of pure GGM
with two scales $\Lambda_{G}$ and $\Lambda_{S}$ at $M_{\rm mess}$ this approach was followed in
\cite{ADJKpGGM,ADJK7}.

One of the main points we want to address in this paper is how
the predictions of GGM, and specifically the consequences of the sfermion mass equations \eqref{scalarsoft}, are affected at lower scales.

The GGM predictions can also be viewed as two constraints (for each of the three generations) coming from the
sfermion mass equation \eqref{scalarsoft}. Indeed there are two matter doublets, $Q$ and $L$, and three singlets,
$U$, $D$ and $E$, but only three scalar mass scales, $\Lambda_{S,r}$, to parameterise them. These two constraints were
expressed by the authors of~\cite{GGM} in terms of the two mass sum rules for each of the three generations,
\be
\Tr\, Y\,m^2 \, =\, 0\,=\, \Tr\, (B-L)\, m^2\, ,
\label{sumrs1}
\ee
or equivalently,
\bea
m_Q^2 -2m_U^2+m_D^2-m_L^2+m_E^2 &=& 0 \, , \label{msr21} \\
2m_Q^2 -m_U^2-m_D^2-2m_L^2+m_E^2 &=& 0 \, . \label{msr22}
\eea
These sum rules have also been obtained and studied earlier in the context of a variety of realisations of SUSY-breaking in \cite{Cohen:2006qc,Martin:1993ft,Faraggi:1991bb,Kawamura:1993uf,Martin:1996zb,Dimopoulos:1996yq}.

In \cite{GGM} it was argued, based on leading order RG equations,
that the sum rules \eqref{msr21}, \eqref{msr22} should
be valid
to high accuracy also at low scales, at least for the first two generations.
Below we will perform the RG evolution numerically using \texttt{SoftSUSY 3.1.6}~\cite{Allanach:2001kg}
and will quantify the status of the sum rules for all three generations.

The strict definition of GGM given above leads to a negligibly small input
value of the soft parameter $B_\mu$ (at the high scale) which consequentially predicts a large value of
$\tan \beta$ at the low (electroweak) scale. The phenomenology of such models (referred to as pure GGM)
was studied in \cite{ADJKpGGM,ADJK7}. In the present paper we relax
this strict implementation of GGM and similarly to \cite{GGM,Komargodski:2008ax}
allow additional non-gauge couplings between the messengers and the
MSSM Higgs fields. These generate $B_\mu$ at the messenger scale which feeds into $\tan \beta$ at the electroweak scale.
In this set-up $\tan \beta$ at the low scale is treated as an independent input parameter of the model
(rather than a prediction as it was the case in the strict GGM, see \cite{ADJKpGGM} for more detail and references).

To summarise the discussion above, we can think of two alternative realisations of the gauge mediation
models parameter space. The first is in terms of $\Lambda_{G,r}$ and $\Lambda_{S,r}$ computed at the high
scale $M_{\rm mess}$, the messenger scale $M_{\rm mess}$ itself, and the value of $\tan \beta$ defined at the electroweak scale.

An alternative proposal is to define the GGM parameter space by constraining the MSSM sfermion
masses (or more precisely the appropriate soft parameters) by the two mass sum rules
\eqref{msr21}, \eqref{msr22} of \cite{GGM}. In this approach the messenger mass scale $M_{\rm mess}$
appears to play no role. In what follows we will investigate the validity of the sum rules at the low scale,
for light and heavy generations, and the role played by $M_{\rm mess}$.

Note that there is a similarity between the mass sum rules of \cite{GGM} and
analytic expressions for one-loop RG invariants of MSSM discussed recently in
\cite{Carena:2010gr,Carena:2010wv}. In a companion paper \cite{JKWrgi} we will investigate GGM in terms of these invariants and further
comment on the messenger scale.

The paper is organised as follows.
In Section \ref{sec:two} we show how the MSSM mass spectrum at the low scale varies as a function
of the messenger scale. We then proceed to
compute the sum rules directly in terms of this mass spectrum
and show their significant violation for the third generation. This direct use of observable masses in the sum rules
neglects the effects of electroweak symmetry breaking on the masses and mixings of sfermions. These effects are taken into
account in Section \ref{sec:three} where we find that the validity of the $B-L$ sum rule for the third generation
is improved, but the corresponding hypercharge sum rule remains broken at a level of up to $50 \%$. We
illustrate that this occurs generically in GGM.
For completeness,
in Section \ref{sec:four} we briefly examine models with more than one messenger scale (two for simplicity,
$M^{\rm high}_{\rm mess}$ and $M^{\rm low}_{\rm mess}$) and find
that not suprisingly the spectrum also shows sensitivity to the intermediate scale $M^{\rm low}_{\rm mess}$.

\begin{figure}[ht!]
\begin{center}
\scalebox{1.03}[1.03]{
\begin{picture}(500,300)(0,0)
\includegraphics[width=0.95\textwidth]{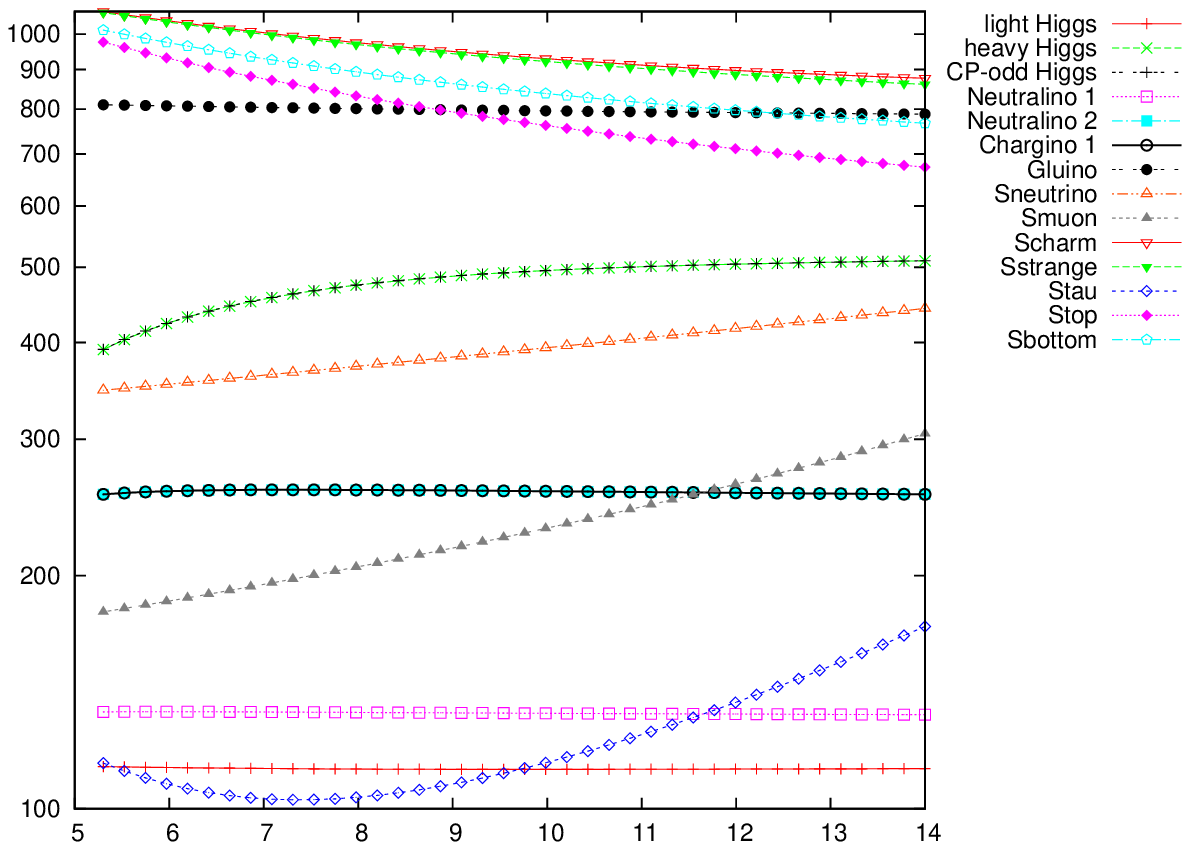}
\rText(-430,165)[c][l]{\scalebox{1}[1]{Mass in GeV}}
\Text(-250,-10)[c]{\scalebox{1}[1]{$\log_{10} \left(M_{\rm mess}/{\rm GeV}\right)$}}
\end{picture}}
\vspace*{0.5cm}
\end{center}
\begin{center}
\caption{
Dependence of the MSSM spectrum on the messenger scale $M_{\rm mess}$ in a simple gauge mediation
model with $\Lambda_{G}(M_{\rm mess})=\Lambda_{S}(M_{\rm mess})=10^5\,{\rm GeV}$ and $\tan \beta = 45$.}
\label{fig:one}
\end{center}
\end{figure}

\section{GGM sum rules in terms of observable masses}\label{sec:two}

Before we turn to the sum rules it is useful to recall that the messenger scale
plays a role and affects phenomenology in the standard approach. In Fig.~\ref{fig:one}
we plot the MSSM spectrum computed at the low scale as a function of the high scale, $M_{\rm mess}$.
The soft parameters $\Lambda_{G,r}$ and $\Lambda_{S,r}$ are introduced at $M_{\rm mess}$ where we start
the RG evolution down to the low scale, and are kept fixed as we vary the starting point $M_{\rm mess}$.
For simplicity we chose all $\L$'s equal (ordinary gauge mediation),
$\Lambda_{G}(M_{\rm mess})=\Lambda_{S}(M_{\rm mess})=10^5\,{\rm GeV}$ and set $\tan \beta = 45$.
As can be seen from Fig.~\ref{fig:one}, the sparticle spectrum changes significantly
when $M_{\rm mess}$ is varied and even the NLSP species can change between neutralino and stau.

One can also check that for different (and lower) values of $\tan \beta$ the variation of
$M_{\rm mess}$ amounts to changes in sparticle masses of similar size as in Fig.~\ref{fig:one}.

This dependence on the messenger scale is even more pronounced in the pure GGM set-up
studied in \cite{ADJKpGGM,ADJK7}. In this case even the allowed region in parameter space
was shrinking as $M_{\rm mess}$ was reduced. This was caused in part by the necessity to generate
phenomenologically viable values of $B_\mu$ through the RG evolution between $M_{\rm mess}$ and the low scale.

Let us now turn to the mass sum rules \eqref{msr21} and \eqref{msr22}. In principle, the variation
of sparticle masses shown in Fig.~\ref{fig:one} does not preclude the validity of the sum rules,
as the effects of individual mass variations can cancel (as will be seen momentarily for the first two
generations).

\begin{figure}
\begin{center}
\hspace*{0.3cm}
\subfigure{\scalebox{0.95}[0.95]{\begin{picture}(500,270)(-30,0)
\includegraphics[width=0.8\textwidth]{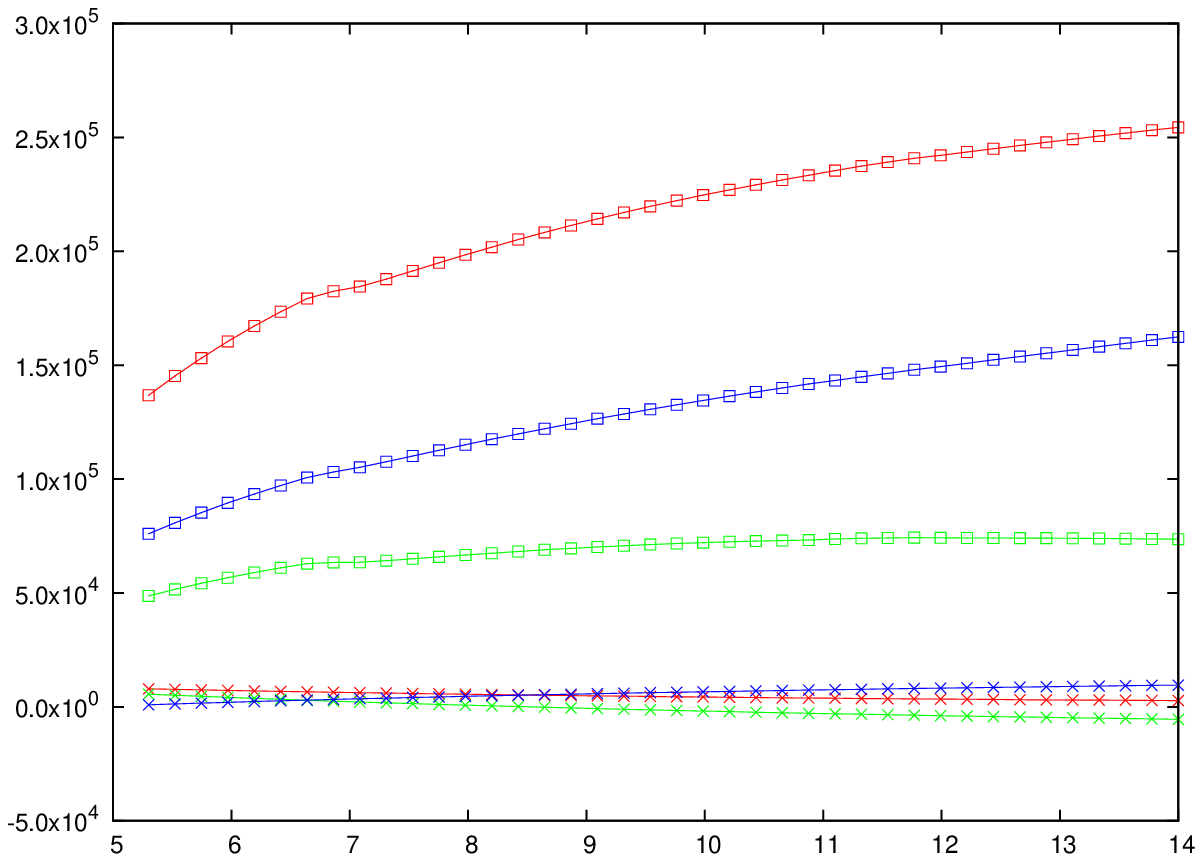}
\rText(-370,135)[c][l]{\scalebox{1}[1]{Sum Rule in GeV$^2$}}
\Text(-167,-10)[c]{\scalebox{1}[1]{$\log_{10} \left(M_{\rm mess}/{\rm GeV}\right)$}}
\end{picture}}}
\vspace*{2.3cm}
\hspace*{0.3cm}
\subfigure{\scalebox{0.95}[0.95]{\begin{picture}(500,270)(-30,0)
\includegraphics[width=0.8\textwidth]{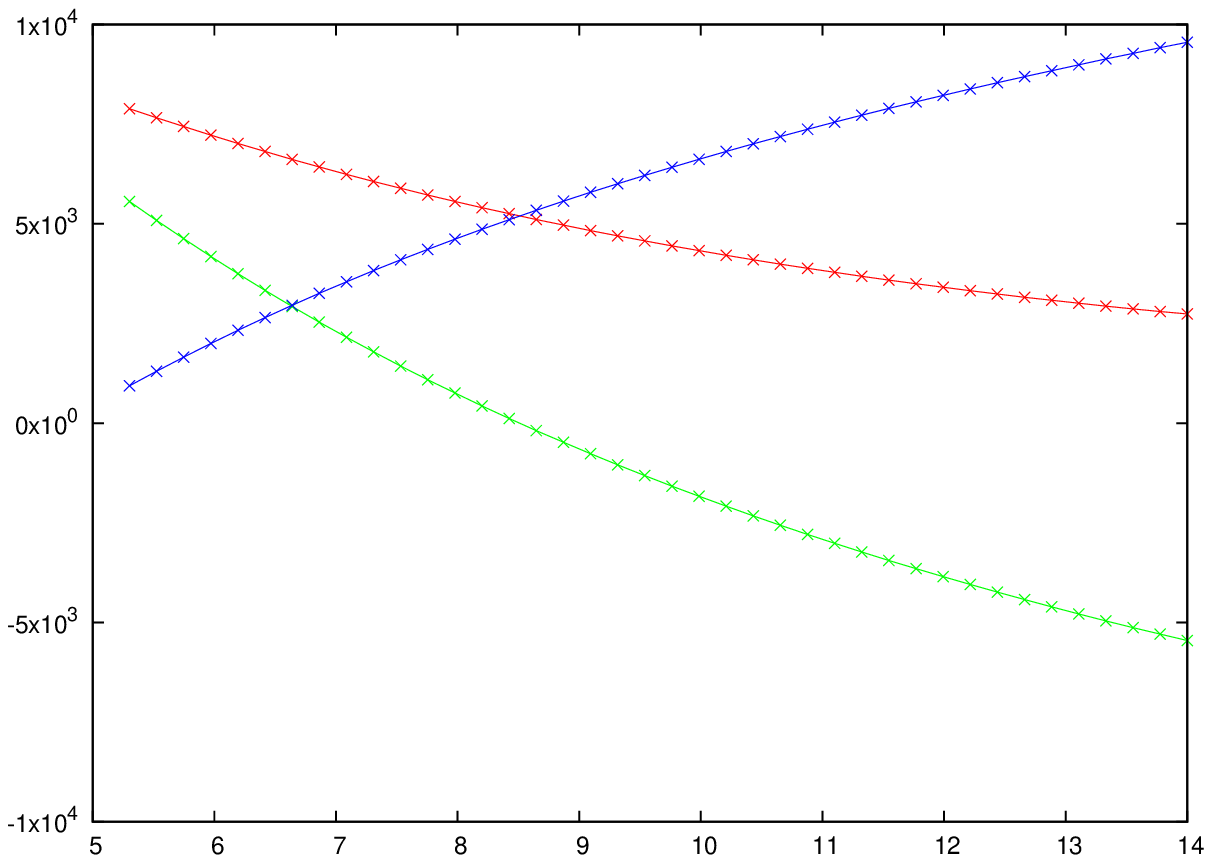}
\rText(-370,135)[c][l]{\scalebox{1}[1]{Sum Rule in GeV$^2$}}
\Text(-167,-10)[c]{\scalebox{1}[1]{$\log_{10} \left(M_{\rm mess}/{\rm GeV}\right)$}}
\end{picture}}}
\end{center}
\vspace*{-2.5cm}
\begin{center}
\caption{Sum rules for mass eigenstates as in~\eqref{naivesr}.
The values of the sum rules in
Eqs.~\eqref{msr21}, \eqref{msr22} and \eqref{sumrs3} are plotted in red, green
and blue respectively versus the messenger scale $M_{\rm mess}$.
Squares indicate the third generation and crosses refer to the
first and second generations. As before we use a simple gauge
mediation model with
$\Lambda_{G}(M_{\rm mess})=\Lambda_{S}(M_{\rm mess})=10^5\,{\rm GeV}$ and
$\tan \beta = 45$.
The lower panel is a blow-up of the upper panel around the origin $\pm 10^4\, {\rm GeV}^2$.
} \label{fig:two}
\end{center}
\end{figure}

To quantify the validity of the sum rules and to elucidate the role of $M_{\rm mess}$, in Fig.~\ref{fig:two}
we plot the right hand sides of Eqs.~\eqref{msr21} and \eqref{msr22} as we vary the messenger scale.
As before, we keep the soft parameters fixed,
$\Lambda_{G}(M_{\rm mess})=\Lambda_{S}(M_{\rm mess})=10^5\,{\rm GeV}$ and take $\tan \beta = 45$.
For now we will evaluate the sum rules by plugging in the values of the observable sfermion masses
in  Eqs.~\eqref{msr21} and \eqref{msr22}.

More precisely, taking for example for the third generation stop, sbottom, stau and sneutrino mass eigenstates\footnote{Where
as usual $m_{\tilde{t}_1}<m_{\tilde{t}_2}$, $m_{\tilde{b}_1}<m_{\tilde{b}_2}$ and $m_{\tilde{\tau}_1}<m_{\tilde{\tau}_2}$.}
$m_{\tilde{t}_1}$, $m_{\tilde{t}_2}$, $m_{\tilde{b}_1}$, $m_{\tilde{b}_2}$, $m_{\tilde{\tau}_1}$, $m_{\tilde{\tau}_2}$ and $m_{\tilde{\nu}_{\tau}}$ we
make an identification (all other matter fields are treated analogously):
\bea
\label{naivesr}
m_U^2&=&m_{\tilde{t}_1}^2 \ , \quad
m_D^2=m_{\tilde{b}_1}^2 \ , \quad
m_Q^2=\frac{1}{2}(m_{\tilde{t}_2}^2 +m_{\tilde{b}_2}^2)\,.
\\\nonumber
m_E^2&=& m^{2}_{\tilde{\tau}_{1}} \ , \quad m^{2}_{L}=\frac{1}{2}(m_{\tilde{\tau}_2}^2 +m^{2}_{\tilde{\nu}_{\tau}})
\eea
This procedure is a very good approximation for the first and second generation,
but ignores sizable mixing effects caused by the large Yukawas and electroweak
symmetry breaking for the third generation. In the following Section we will carefully account for this.
For now we will stick to this simplified procedure dictated by \eqref{naivesr} as it appeals only to
the experimentally observable masses.

The two sum rules \eqref{msr21} and \eqref{msr22} along with a linear combination
of them discussed in \cite{GGM},
\be
\Tr\, Y\,m^2 \, -\, \frac{5}{4}\,\Tr\, (B-L)\, m^2\,=\, 0\, ,
\label{sumrs3}
\ee
are plotted in Fig.~\ref{fig:two} for the third and for the first generation (the latter
being indistinguishable from the second generation). For the sum rules to hold, their
plotted value should be much smaller than a typical mass-squared contribution. In our case the
latter are of the order $2\times 10^6\, {\rm GeV}^2$. The figure shows that the sum rules for the first
and second generations are satisfied with an accuracy of $\lesssim 1\%$. However, both the $B-L$, and the $Y$
sum rules for the third generation are violated by $\sim 3\%$ and $\sim 15\%$, respectively. We note that
these sum rules apart from being non-vanishing, also vary with $M_{\rm mess}$.

For more general models with distinct $\Lambda_{G,r}$ and $\Lambda_{S,r}$ we found that the $(B-L)$ and hypercharge sum rules are violated by
up to $6\%$ and $60\%$, respectively.  (We have  allowed for an order of magnitude split between $\Lambda_{S}$ and $\Lambda_{G}$ in both directions, allowed for different values of the $\Lambda_{r}$ and have also varied $\tan\beta$ between $2$ and $45$.)

It transpires that the sum rules for the
first two generations are indeed a prediction of GGM which can in principle be directly accessed at the collider scale.
But in order to do so, one has to be able to measure and identify all the superpartner species
and distinguish the first two from the third generation. The latter sparticles
are typically significantly lighter due to large Yukawas.

The third generation sum rules have failed chiefly due to effects of the large Yukawa interactions.
In the following Section we will check if the situation for the third generation can be improved by
using the more appropriate soft mass terms rather than the mass eigenstates.

\section{GGM sum rules in terms of soft masses}\label{sec:three}

Electroweak symmetry breaking contributes to the masses of the sfermions and induces mixing between the left-
and right-handed sfermions of a given flavour. In the previous Section we have expressed the sum rules
in terms of the mass eigenstates through a simple identification \eqref{naivesr} which ignored these effects.
In reality, the sum rules arise at the high scale where they are written in terms of the soft SUSY-breaking
masses, $m_Q^2$, $m_U^2$, etc and not in terms of the sfermion mass eigentates $m_{\tilde{u}_{1}}^{2}$, $m_{\tilde{u}_{2}}^{2}$,
and so on. In this Section we will compute the sum rules in terms of the soft masses.
The soft masses at the low scale can be extracted from the observable mass eigenstates (which we read off
\texttt{SoftSUSY}) as follows.

Diagonalisation of the mass matrix gives the mass eigenstates
and defines a mixing angle (see e.g. \cite{Martin:1997ns}). At tree-level, one has for the up-type squarks,
\bea
 \label{eq:diagonalisation}
\begin{pmatrix}
m_{\tilde{u}_{i,1}}^{2} & 0
\\ 0 & m_{\tilde{u}_{i,2}}^{2}
\end{pmatrix} &=&
\begin{pmatrix}
\cos\theta_{\tilde{u}_{i}} & -\sin\theta_{\tilde{u}_{i}}
\\ \sin\theta_{\tilde{u}_{i}} & \cos\theta_{\tilde{u}_{i}}
\end{pmatrix}
M_{\tilde{u}_{i}}^{2}
\begin{pmatrix}
\cos\theta_{\tilde{u}_{i}} & \sin\theta_{\tilde{u}_{i}}
\\ -\sin\theta_{\tilde{u}_{i}} & \cos\theta_{\tilde{u}_{i}}
\end{pmatrix}\, ,
\\ \nonumber \\
 \label{eq:upsquarkmixingMat}
M_{\tilde{u}_{i}}^{2} &=&
\begin{pmatrix}
(m_{Q}^{2})_{ii} + m_{u,i}^{2} + \Delta_{\tilde{u}_{L}} & m_{u_{i}}\left((A_{u})_{ii} - \mu \cot \beta \right)
\\ m_{u_{i}}\left((A_{u})_{ii} - \mu \cot \beta \right) & (m_{U}^{2})_{ii} + m_{u,i}^{2} + \Delta_{\tilde{u}_{R}}
\end{pmatrix}\, ,\\ \nonumber \\
\Delta_{\tilde{f}} &\equiv & (T_{3,\tilde{f}} \, - \, Q_{\tilde{f}}\sin^{2}\theta_{W}) \cos(2\beta) m_{Z}^{2}\, .
 \label{eq:upsquarkmixing}
\eea
where the left hand side of \eqref{eq:diagonalisation} gives the observable mass eigenstates calculated by \texttt{SoftSUSY}.
Equation \eqref{eq:upsquarkmixingMat} gives the original mass matrix in the
$(\tilde{u}_{i,L}, \tilde{u}_{i,R})^{T}$ basis. There we are after the soft masses, given by
 $(m_{Q}^{2})_{ii}$ and $(m_{U}^{2})_{ii}$, for the $i$th generation
of the left-handed doublet and the right-handed up-type squarks.
$(A_{u})_{ij}$ is the matrix of up-type soft $A$-parameters -- the scalar trilinear
couplings divided by the corresponding Yukawa couplings. $T_{3,\tilde{f}}$ and $Q_{\tilde{f}}$ are the third component of weak
isospin and the electric charge.
\begin{figure}
\begin{center}
\hspace*{0.3cm}
\subfigure{\scalebox{0.95}[0.95]{\begin{picture}(500,270)(-30,0)
\includegraphics[width=0.8\textwidth]{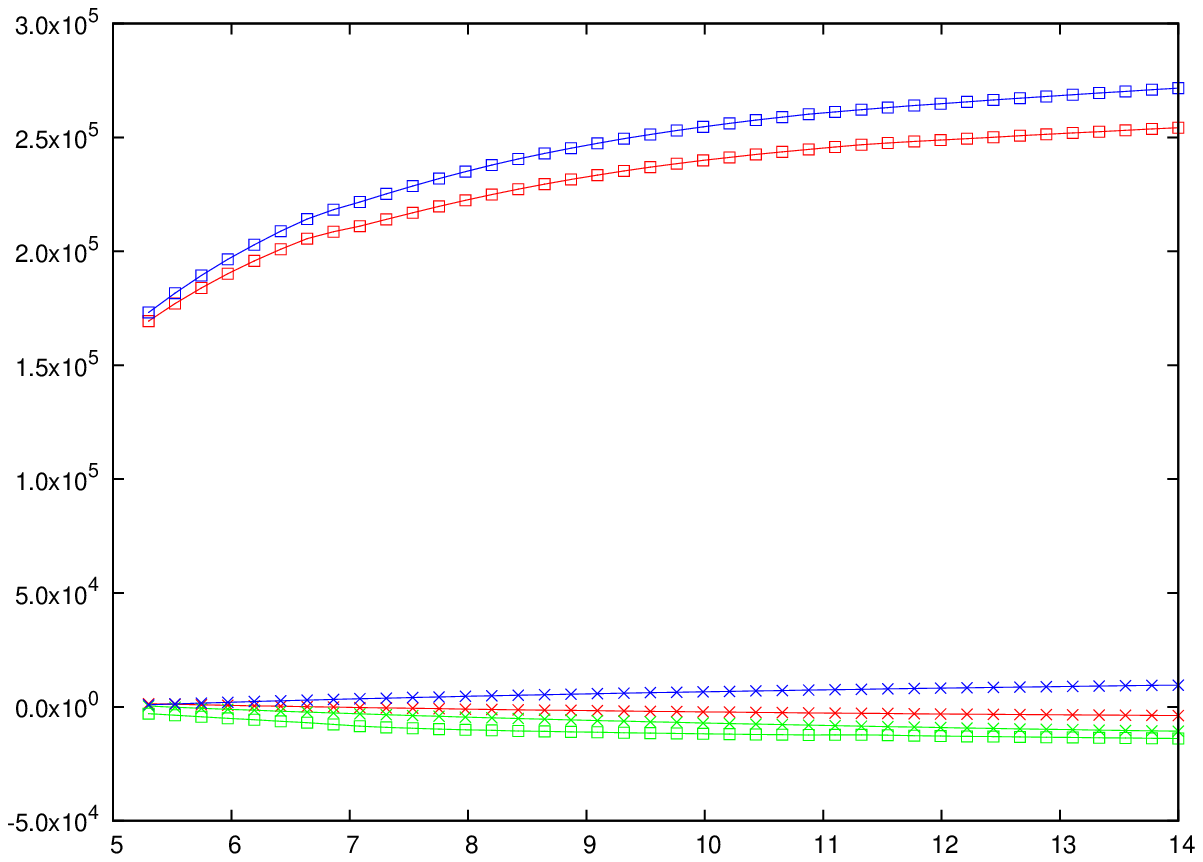}
\rText(-370,135)[c][l]{\scalebox{1}[1]{Sum Rule in GeV$^2$}}
\Text(-167,-10)[c]{\scalebox{1}[1]{$\log_{10} \left(M_{\rm mess}/{\rm
GeV}\right)$}}
\end{picture}}}
\vspace*{2.3cm} \hspace*{0.3cm}
\subfigure{\scalebox{0.95}[0.95]{\begin{picture}(500,270)(-30,0)
\includegraphics[width=0.8\textwidth]{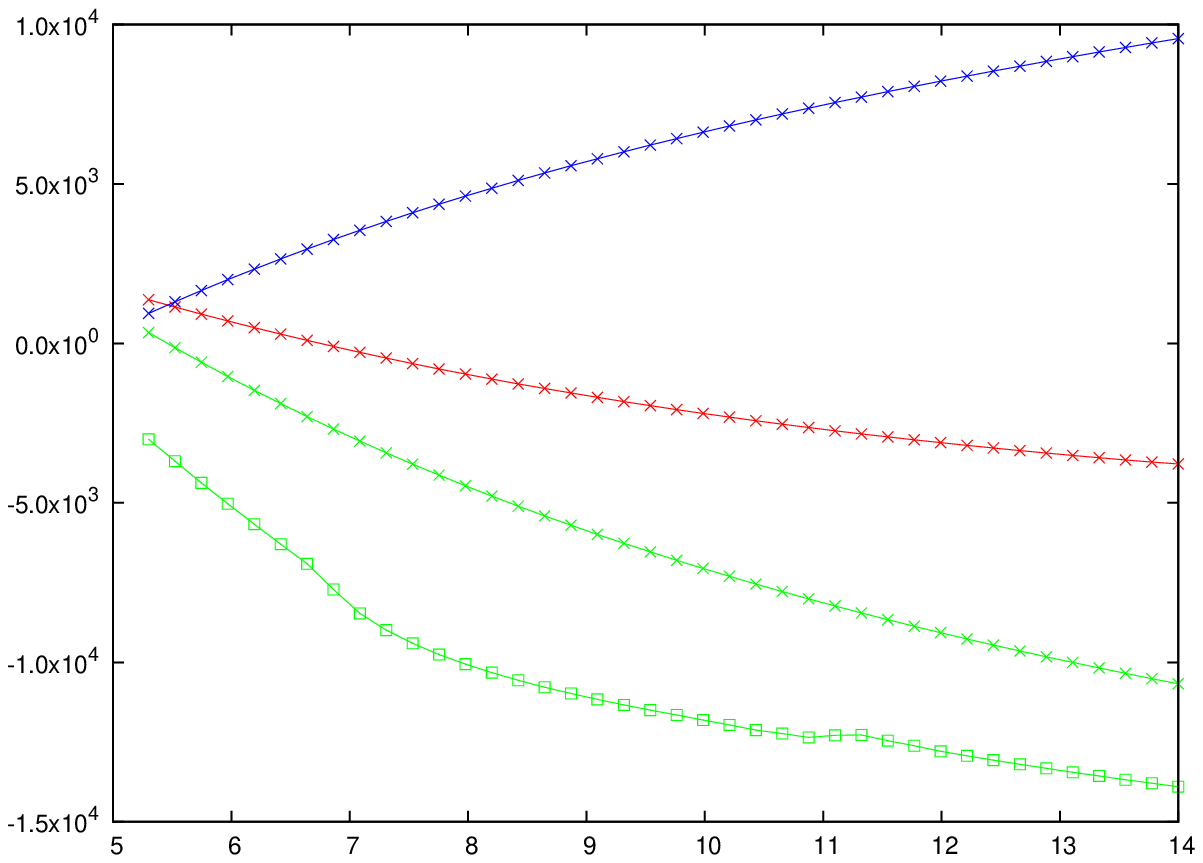}
\rText(-370,135)[c][l]{\scalebox{1}[1]{Sum Rule in GeV$^2$}}
\Text(-167,-10)[c]{\scalebox{1}[1]{$\log_{10} \left(M_{\rm mess}/{\rm
GeV}\right)$}}
\end{picture}}}
\end{center}
\vspace*{-2.5cm}
\begin{center}
\caption{Soft-mass sum rules for a high $\tan \beta$ OGM model.
Equations~\eqref{msr21},
\eqref{msr22} and \eqref{sumrs3} are plotted in red, green and blue
respectively versus the messenger scale $M_{\rm mess}$. Squares
indicate the third generation and crosses refer to the first and
second generations. We use a simple gauge mediation
model with $\Lambda_{G}(M_{\rm mess})=\Lambda_{S}(M_{\rm mess})=10^5\,{\rm
GeV}$ and $\tan \beta = 45$. The lower panel is a zoom of the upper panel.} \label{fig:three}
\end{center}
\end{figure}

\begin{figure}
\begin{center}
\hspace*{0.3cm}
\subfigure{\scalebox{0.95}[0.95]{\begin{picture}(500,270)(-30,0)
\includegraphics[width=0.8\textwidth]{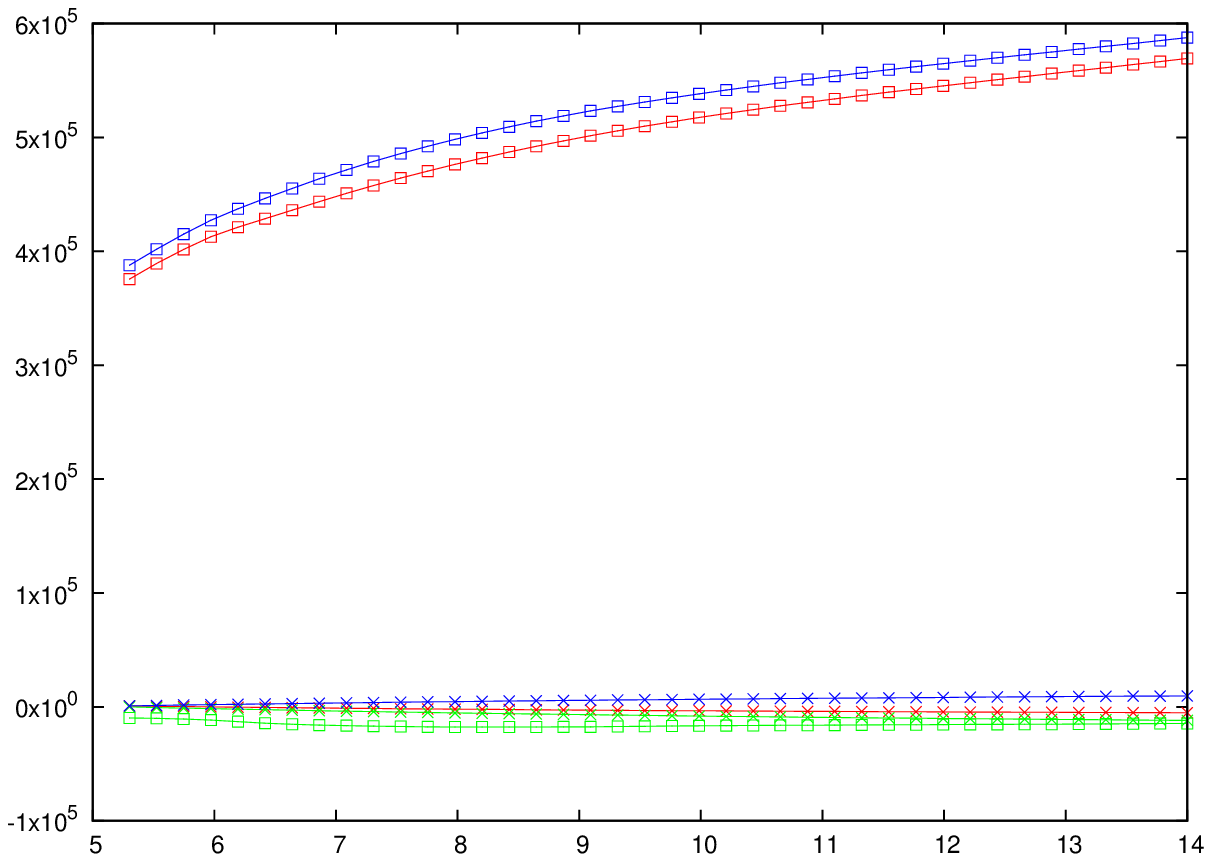}
\rText(-370,135)[c][l]{\scalebox{1}[1]{Sum Rule in GeV$^2$}}
\Text(-167,-10)[c]{\scalebox{1}[1]{$\log_{10} \left(M_{\rm mess}/{\rm
GeV}\right)$}}
\end{picture}}}
\vspace*{2.3cm} \hspace*{0.3cm}
\subfigure{\scalebox{0.95}[0.95]{\begin{picture}(500,270)(-30,0)
\includegraphics[width=0.8\textwidth]{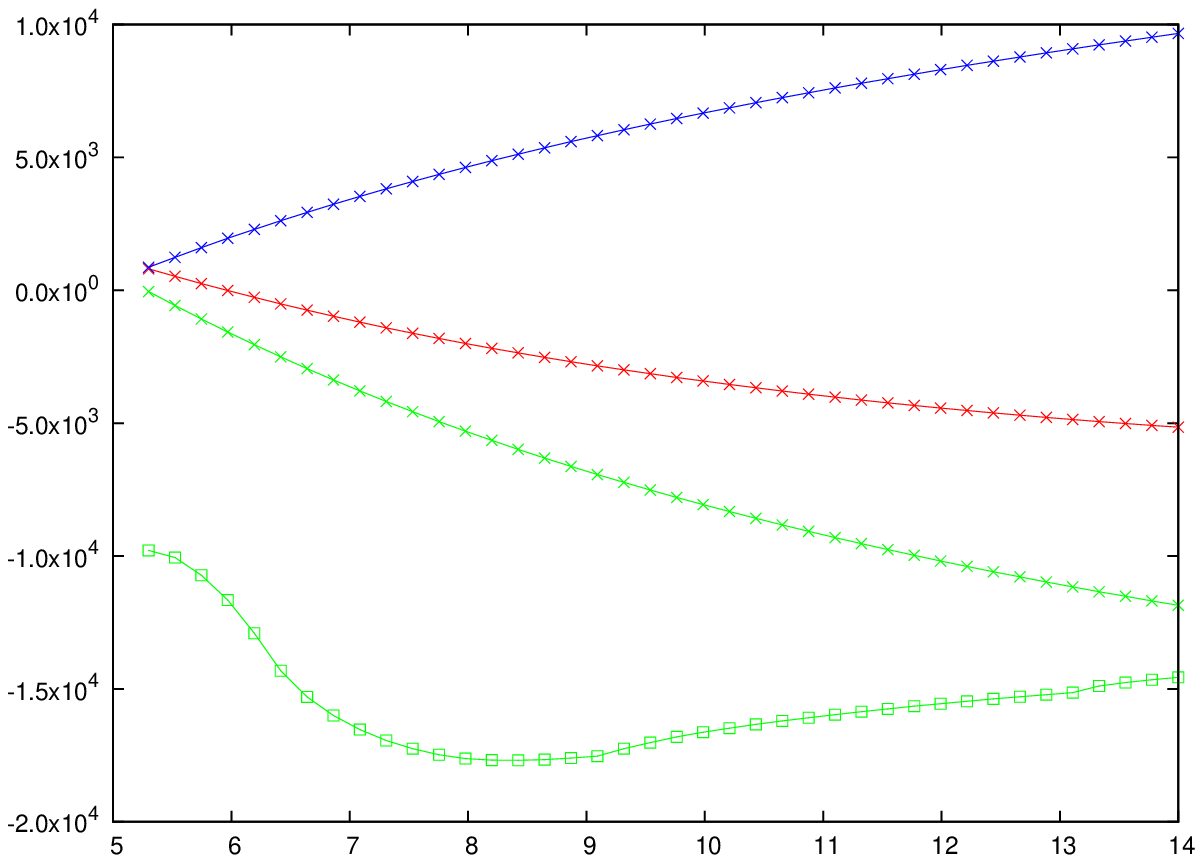}
\rText(-370,135)[c][l]{\scalebox{1}[1]{Sum Rule in GeV$^2$}}
\Text(-167,-10)[c]{\scalebox{1}[1]{$\log_{10} \left(M_{\rm mess}/{\rm
GeV}\right)$}}
\end{picture}}}
\end{center}
\vspace*{-2.5cm}
\begin{center}
\caption{Soft-mass sum rules for a low $\tan \beta$ OGM model.
Equations~\eqref{msr21},
\eqref{msr22} and \eqref{sumrs3} are plotted in red, green and blue
respectively versus the messenger scale $M_{\rm mess}$. Squares
indicate the third generation and crosses refer to the first and
second generations. We use a simple gauge mediation
model with $\Lambda_{G}(M_{\rm mess})=\Lambda_{S}(M_{\rm mess})=10^5\,{\rm
GeV}$ and $\tan \beta = 2$. The lower panel is a zoom of the upper panel.} \label{fig:three-low}
\end{center}
\end{figure}

The off-diagonal terms in \eqref{eq:upsquarkmixingMat}
are proportional
to the relevant fermion mass which, compared to the soft mass parameters, is negligible for the first two generations
(indeed \texttt{SoftSUSY} sets these terms to zero). The mass eigenstates then align with the $L/R$ eigenstates\footnote{Note
that while the soft $A$-parameters contain a division by the Yukawa couplings, the assumption of soft
supersymmetry-breaking universality takes the scalar trilinear couplings to be proportional to the corresponding Yukawa
couplings, so that the $L$-$R$ mixing terms in \eqref{eq:upsquarkmixingMat} should indeed be negligible for the light generations.}.

The analogue of these equations for the
down-type quarks is obtained by substituting
$ u, \cot\beta \rightarrow d, \tan\beta$. Similarly for the charged sleptons one has
 $ Q, u, \cot\beta \rightarrow L, e, \tan\beta$.

To extract the soft mass parameters, we simply need to rotate the mass eignestaters back through the appropriate mixing angle
 and to subtract the known electroweak contribution
$m_{f,i}^{2} + \Delta_{\tilde{f},L/R}$ ($\Delta_{\tilde{f}} \equiv \Delta_{f}$)
from the $LL/RR$ elements of the resulting mass matrices.
The mixing angle,
$\theta_{\tilde{t}}$, $\theta_{\tilde{b}}$ and $\theta_{\tilde{\tau}}$, is non-zero only for the third generation,
and in principle can be determined experimentally from electroweak coupling strengths (for us we can read it from
\texttt{SoftSUSY}).
In this way, $(m_{Q}^{2})_{ii}$ is given by both
$(M_{\tilde{u}_{i}}^{2})_{LL}$ and $(M_{\tilde{d}_{i}}^{2})_{LL}$; $(m_{L}^{2})_{ii}$ is also determined both by
$(M_{\tilde{e}_{i}}^{2})_{LL}$ and the $i$th generation sneutrino mass
$m_{\tilde{\nu}_{i}}^{2} = (m_{L}^{2})_{ii} + \Delta_{\tilde{\nu}}$. The two corresponding values always agreed to better than
1\% (and better still for the lighter generations).

The two sum rules \eqref{msr21} and \eqref{msr22} along with the linear combination
\eqref{sumrs3}
are plotted in Fig.~\ref{fig:three} for a simple gauge mediation model with a high value of
$\tan \beta= 45$. The same model with a low value of $\tan \beta= 2$ is shown in Fig.~\ref{fig:three-low}.
From these figures one can see that the $B-L$ mass sum rule for the third generation is now improved, and
holds to better than $1 \%$ accuracy. At the same time, the third generation hypercharge sum rule remains
broken. It is violated by $10$ to $20 \%$ in the high $\tan \beta$ OGM model and by
$20$ to $45 \%$ in the low $\tan \beta$ case. We have also computed sum rules for non-OGM models
with non-equal $\L_G$ and $\L_S$ parameters. In these cases we found violation of the third generation
hypercharge sum rule by up to $60 \%$.

\section{Comments on multiple messenger scales}\label{sec:four}

So far we have been discussing GGM models with a single messenger scale.
All of our findings can be straightforwardly generalised for models with
more than one messenger scale. To illustrate this point we will consider here
a simple class of models with two messenger scales, $M_{\rm mess}^{\rm high}$ and $M_{\rm mess}^{\rm low}$.

Specifically, we will look at the variation of the sparticle spectrum keeping the
high scale, $M_{\rm mess}^{\rm high}$, fixed and changing the intermediate scale, $M_{\rm mess}^{\rm low}$.
We define effective $\L$-parameters,
\bea
\label{effective}
\L_G^{\rm eff} &=& \L_G^{\rm low} +\L_G^{\rm high} \, , \\\nonumber
(\L_S^{\rm eff})^2 &=& (\L_S^{\rm low})^2 +(\L_S^{\rm high})^2 \, ,
\eea
which we keep fixed while varying $M_{\rm mess}^{\rm low}$.
The individual contributions $\L^{\rm low}$ and $\L^{\rm high}$ arise from messengers being
integrated out at the corresponding low and high scale.

\begin{figure}
\begin{center}
\hspace*{0.3cm}
\subfigure{\scalebox{0.95}[0.95]{\begin{picture}(500,270)(-30,0)
\includegraphics[width=0.8\textwidth]{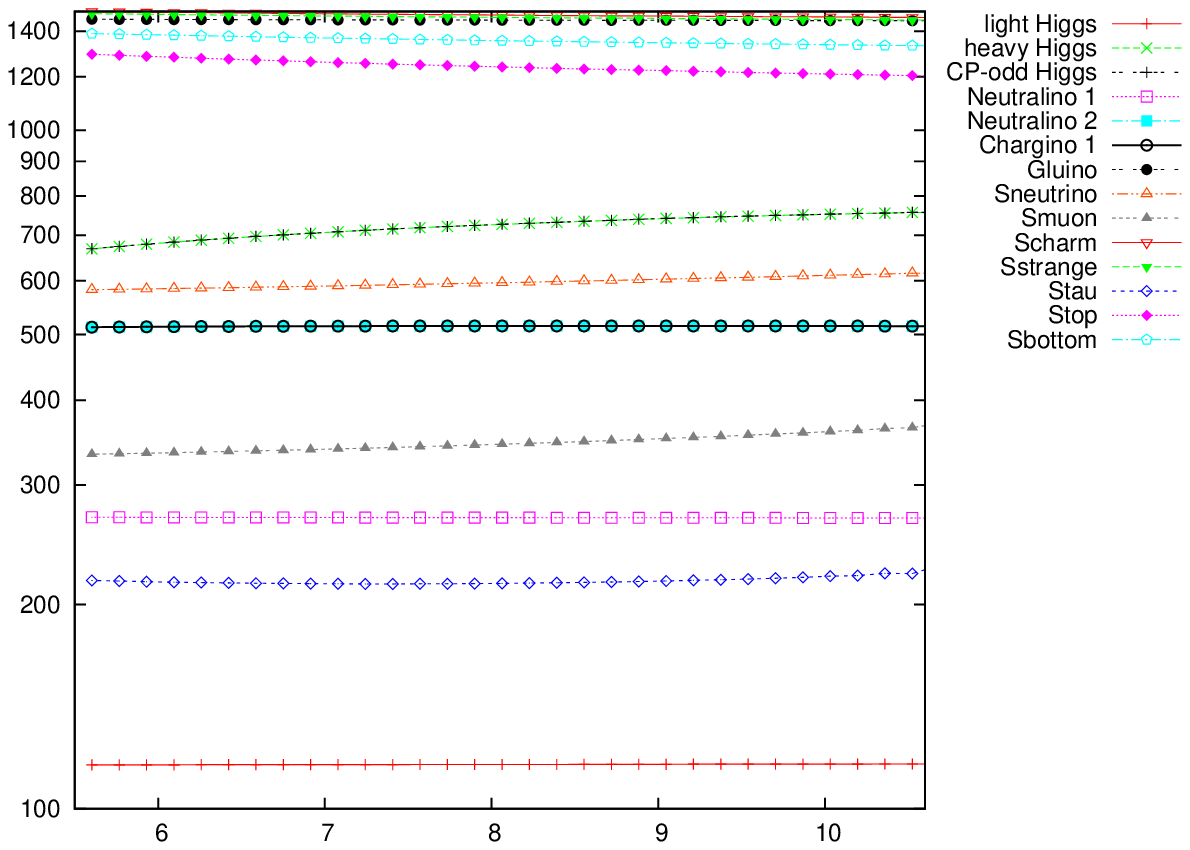}
\rText(-370,135)[c][l]{\scalebox{1}[1]{Mass in GeV}}
\Text(-205,-10)[c]{\scalebox{1}[1]{$\log_{10} \left(M^{\rm low}_{\rm mess}/{\rm
GeV}\right)$}}
\end{picture}}}
\vspace*{2.3cm} \hspace*{0.3cm}
\subfigure{\scalebox{0.95}[0.95]{\begin{picture}(500,270)(-30,0)
\includegraphics[width=0.8\textwidth]{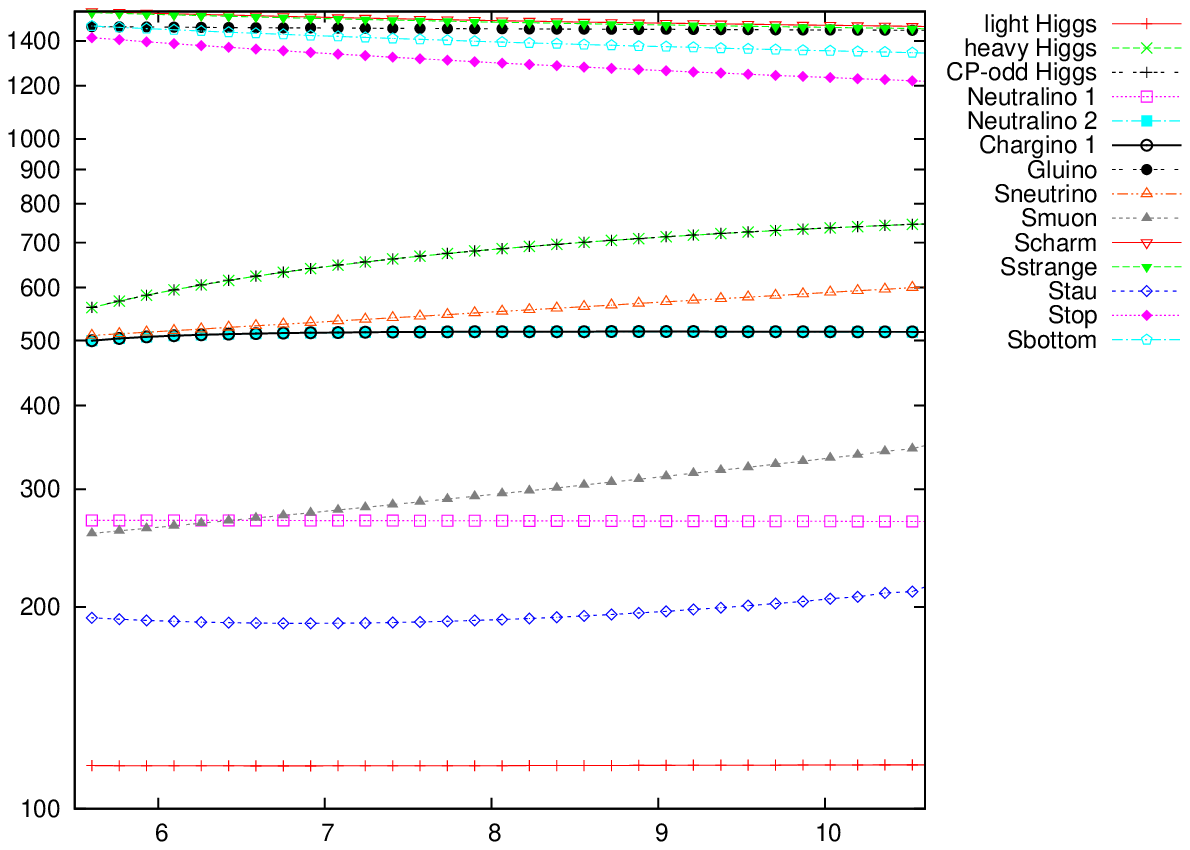}
\rText(-370,135)[c][l]{\scalebox{1}[1]{Mass in GeV}}
\Text(-205,-10)[c]{\scalebox{1}[1]{$\log_{10} \left(M^{\rm low}_{\rm mess}/{\rm
GeV}\right)$}}
\end{picture}}}
\end{center}
\vspace*{-2.5cm}
\begin{center}
\caption{Physical mass spectrum for two setups with two messenger scales. The contributions to $\Lambda_{G,S}$
from messengers at scales $M^{\rm high/low}_{\rm mess}$ are given by Eq.~\eqref{splitequal} (upper panel) and
by Eq.~\eqref{splitnequal} (lower panel).
We keep $\Lambda^{\rm eff}_{G}=2\times 10^5\,{\rm GeV}$, $(\Lambda^{\rm eff})^{2}=2\times (10^{5}\,{\rm GeV})^2$ and
$M^{\rm high}_{\rm mess}=10^{12} \,\rm{GeV}$ fixed as we vary the
intermediate messenger scale $M^{\rm low}_{\rm mess}$.
$\tan \beta = 45$.} \label{fig:five}
\end{center}
\end{figure}

We consider two different examples for the relative sizes of $\L^{\rm low}$ and $\L^{\rm high}$.
In the first case they are split equally, $\L^{\rm low}=\L^{\rm high}$. This example is
motivated by the simplest ordinary gauge mediation set-up where each $\L$ behaves as
\be
\L^{\rm low/high} \sim \frac{\lambda_{\rm mess}^{\rm low/high} F_{\Phi}}{M_{\rm mess}^{\rm low/high}} \, =\,
\frac{\lambda_{\rm mess}^{\rm low/high} F_{\Phi}}{\lambda_{\rm mess}^{\rm low/high}\langle \Phi \rangle}
\, =\,\frac{F_{\Phi}}{\langle \Phi \rangle}\, .
\label{splitequal}
\ee
In the Fig.~\ref{fig:five} we show the variation of the spectrum in the MSSM as we vary $M_{\rm mess}^{\rm low}$ with
$M_{\rm mess}^{\rm high}$ fixed at $10^{14}\, {\rm GeV}$. The effective $\L$-scales are fixed at
$\L_G^{\rm eff}= 2\times 10^{5}\, {\rm GeV}$ and $(\L_S^{\rm eff})^2= 2\times (10^{5}\, {\rm GeV})^2$ and
$\tan \beta =45$.
Similarly to the single-scale case of Fig.~\ref{fig:one} we observe a sizable variation of the spectrum.

In the second approach we consider $M_{\rm mess}$ originating from a different source, not directly related to the vev $\langle \Phi\rangle$. More precisely, we vary the balance between $\Lambda^{\rm low}$ and $\Lambda^{\rm high}$ according to
\begin{equation}
\frac{\Lambda^{\rm high}}{\Lambda^{\rm low}} = \frac{\lambda_{\rm mess} F/M^{\rm high}_{\rm mess}}{\lambda_{\rm mess} F/M^{\rm low}_{\rm mess}} = \frac{M^{\rm low}_{\rm mess}}{M^{\rm high}_{\rm mess}}
\label{splitnequal}
\end{equation}
(still holding $\Lambda^{\rm eff}$, defined in Eq.~\eqref{effective}, constant).

For the calculation, we supplemented \texttt{SoftSUSY} with \texttt{HidSecSoftsusy} \cite{KerenZur:2010zy} to permit a second messenger field to dynamically feed into the (until now purely MSSM) RG evolution below the highest scale. At the high scale $M^{\rm high}_{\rm mess}$, the soft masses are detrmined by $\Lambda^{\rm high}_{G,S}$. At the intermediate scale $M^{\rm low}_{\rm mess}$ integrating out the lighter (SU(5) $5 \oplus \overline{5}$) messenger contributes an additional $\Lambda^{\rm low}_{G,S}$ threshold effect to the soft masses.
In between $M^{\rm high}_{\rm mess}$ and $M^{\rm low}_{\rm mess}$, the lighter messenger adds to the beta functions of the gauge couplings and scalar masses at one-loop as described in \cite{KerenZur:2010zy}.
In Fig.~\ref{fig:five} we show the results for $M^{\rm low}_{\rm mess}$ varying over a range not fully extending up to $M^{\rm high}_{\rm mess}$: when the two became too close, \texttt{HidSecSoftsusy} encountered numerical instabilities in calculating the spectrum. (It is worth noting however that when the two scales coincide, $M^{\rm low}_{\rm mess} = M^{\rm high}_{\rm mess}$, the result reproduces that of the corresponding OGM case, calculated with regular \texttt{SoftSUSY}.)  The smaller range of messenger scales used causes the resulting spectra to look flatter, at first glance, when compared to Fig.~\ref{fig:one}. Indeed even taking this into account it is slightly flatter due to the contribution of the second messenger whose mass is
kept fixed (this is less pronounced when the contribution of to the $\Lambda^{\rm eff}$ is distributed according to Eq.~\eqref{splitnequal}
where most of the $\Lambda^{\rm eff}$ can be attributed to the lighter messenger).

We note for completeness that the sum rules are violated by amounts similar to the one scale case.

\section{Conclusions}\label{sec:concl}

In the context of general gauge mediation (GGM) the messenger scale is often taken to be a non-parameter and consequently ignored.
In this paper we have investigated the role of the messenger scale with a particular eye on the MSSM mass spectrum and the validity of sum rules predicted in the GGM framework.

We found that both in the context of models with single and multiple messenger scales, the sparticle mass spectrum depends significantly on the messenger scale(s). Indeed when $M_{\rm mess}$ is varied even the NLSP species can change between neutralino and stau.

Looking at the sum rules we found that the hypercharge sum rule for the third generation is violated by up to $50\%$. This is phenomenologically relevant since sparticles of the third generation are typically the lightest. Furthermore the amount by which this sum role is broken is dependent on the messenger mass.

The remaining (B-L) sum rule for the third generation holds to an accuracy of $\sim 1\%$ if the sum rule is interpreted in terms of soft mass parameters rather than the more directly observable physical masses\footnote{Otherwise, i.e. naively plugging in the observable masses, this sum rule is also broken.}. Importantly all the sum rules for the first two generations hold to a very good accuracy. Therefore, (subject to being able to measure the sparticle mass spectrum and to discriminate between the generations) they could provide a good test for GGM.

\subsection*{Acknowledgements}
We are grateful to Steve Abel and Matt Dolan for useful discussions
and comments, and Boaz Keren-Zur for helpful communications on the use of HidSecSOFTSUSY.


\newpage
\providecommand{\href}[2]{#2}\begingroup\raggedright
\endgroup
\end{document}